\documentclass[fleqn,usenatbib]{mnras}

\usepackage{mathptmx}
\usepackage[T1]{fontenc}
\usepackage{ae,aecompl}

\usepackage{graphicx}	% Including figure files
\usepackage{amsmath}	% Advanced maths commands
\usepackage{amssymb}	% Extra maths symbols

\usepackage{hyperref}
\newcommand{\muhz}{\mbox{$\mu$Hz}}

\newcommand{\numax}{\mbox{$\nu_{\rm max}$}}
\newcommand{\teff}{\mbox{$T_{\rm eff}$}}
\newcommand{\pgran}{\mbox{$P_{\rm gran}$}}
\newcommand{\logg}{\mbox{$\log g$}}
\newcommand{\feh}{\mbox{$\rm{[Fe/H]}$}}
\newcommand{\kp}{\mbox{Kp}}
\newcommand{\kepler}{{\em Kepler}}

\newcommand{\gaia}{\mbox{\textit{Gaia}}}
\newcommand{\lsun}{\ensuremath{\,L_\odot}}

\newcommand{\new}[1]{#1}

\newcommand{\old}[1]{\relax}

\title[Surface gravities for \kepler\ stars]{Surface gravities for
15,000 \kepler\ stars measured from stellar granulation and validated with Gaia DR2 parallaxes}

\author[Pande et al.]{
Durlabh Pande,$^{1,2}$
Timothy R. Bedding,$^{1,2}$\thanks{E-mail: tim.bedding@sydney.edu.au}
Daniel Huber$^{3,1,4,2}$ and
Hans~Kjeldsen$^{2}$
\\
$^{1}$Sydney Institute for Astronomy, School of Physics, University of Sydney 2006, Australia \\
$^{2}$Stellar Astrophysics Centre, Department of Physics and Astronomy, Aarhus University, Denmark \\
$^{3}$Institute for Astronomy, University of Hawai`i, 2680 Woodlawn Drive, Honolulu, HI 96822, USA \\
$^{4}$SETI Institute, 189 Bernardo Avenue, Mountain View, CA 94043, USA}

\date{Accepted XXX. Received YYY; in original form 2018 June 11}

\pubyear{2018}

\begin{document}
\label{firstpage}
\pagerange{\pageref{firstpage}--\pageref{lastpage}}
\maketitle

\begin{abstract}
We have developed a method to estimate surface gravity (\logg) from light curves by measuring the granulation background, similar to the ``flicker'' method by \citet{Bastien2016} but working in the Fourier power spectrum.  We calibrated the method using {\em Kepler} stars for which asteroseismology has been possible with short-cadence data, demonstrating a precision in \logg\ of about 0.05\,dex. We also derived a correction for white noise as a function of {\em Kepler} magnitude by measuring white noise directly from observations. We then applied the method to the same sample of long-cadence stars as \citeauthor{Bastien2016}  \new{We found that about half
the stars are too faint for the granulation background to be reliably detected above the white noise.}  We provide a catalogue of \logg\ values for about 15,000 stars having uncertainties better than 0.5\,dex. We used \gaia\ DR2 parallaxes to validate that granulation is a powerful method to measure \logg\ from light curves. Our method can also be applied to the large number of light curves collected by K2 and TESS.
\end{abstract}


%%%%%%%%%%%%%%%%%%%%%%%%%%%%%%%%%%%%%%%%%%%%%%%%%%

%%%%%%%%%%%%%%%%% BODY OF PAPER %%%%%%%%%%%%%%%%%%

\section{Introduction}\label{sec:intro}

The surface gravity of a star is one of its most fundamental parameters.
It is of particular interest for stars hosting transiting exoplanets, due to its direct dependence on the stellar (and hence planetary) radius. Traditionally, \logg\ is estimated from high-resolution spectra by measuring pressure broadening of absorption lines \new{or using ionization equilibrium}, but \old{this approach is} \new{these approaches are} often plagued by degeneracies with other atmospheric parameters such as temperature and metallicity \citep{Torres2012}. Asteroseismology using data from the {\em Kepler} Mission \citep{Borucki2010} has been demonstrated to provide accurate surface
gravities for the brighter FGK stars \citep{chaplin14,serenelli17}.  For red giants, which oscillate at relatively low frequencies, observations in long-cadence mode (29.4 minutes) can be used \citep[e.g.,][]{Yu2018}.  However, measuring oscillations in main-sequence and subgiant stars requires short-cadence observations (58.9\,s), which were only obtained for a small fraction of \kepler\ stars.  For those tens of thousands of \kepler\ stars that were only observed in long-cadence mode, which includes the majority of
stars hosting transiting planet candidates, it is still possible to infer \logg\ by measuring the low-frequency photometric fluctuations from granulation \citep{mathur11b, Bastien2013, Cranmer2014, kallinger16, Bugnet2017, Ness2018}.  For example, \citet[][hereafter B16]{Bastien2016} measured granulation in the time series using a fixed 8-hour filter, which they referred to as ``flicker" ($F_8$), deriving \logg\ for nearly 28,000 stars with a typical precision of about 0.1 dex. Somewhat controversially, flicker results implied that nearly 50\% of all bright exoplanet host stars are subgiants, resulting in 20\%--30\% larger planet radii than previous measurements had suggested \citep{Bastien2014}. However, radius estimates using the recently-released \gaia\ DR2 parallaxes imply an upper limit on the subgiant fraction of $\sim$\,23\% \citep{berger18}.

A common approach for previous methods using long-cadence data is to measure granulation power over a fixed frequency range, either in the time domain (B16) or in the frequency domain \citep{Bugnet2017}. Here, we present a new method that adapts to the fact that the granulation timescale shifts in frequency depending on the evolutionary state of the star \citep{kb11}.
Additionally, our method is able to provide reliable \logg\ uncertainties, thus allowing a quantitative estimate of whether granulation has been reliably detected in each star. Both of these developments will be critical to applying granulation-based \logg\ estimators to the large number of stars expected to be observed with the TESS mission.

%-----------------------------------------------------------------
\section{Method}\label{sec:method}

\subsection{Surface gravities from photometric time series}

For stars in which solar-like oscillations can be measured, it is now quite
well-established that the frequency of maximum oscillation power, \numax,
scales in a simple way with surface gravity and effective temperature
\citep{Brown1991,Kjeldsen1995}:
\begin{equation}
  \numax \propto g / \sqrt{\teff}. \label{eq:numax.scaling}
\end{equation}
Using this scaling relation, \logg\ can be measured from asteroseismology
with an accuracy of about 0.03 dex for main-sequence stars, based on comparisons with oscillating stars in binary systems \citep{miglio11,huber14b}.

\begin{figure}
\begin{center}
\includegraphics[width=0.49\textwidth, trim={0cm 0cm 1.5cm 0cm},clip]{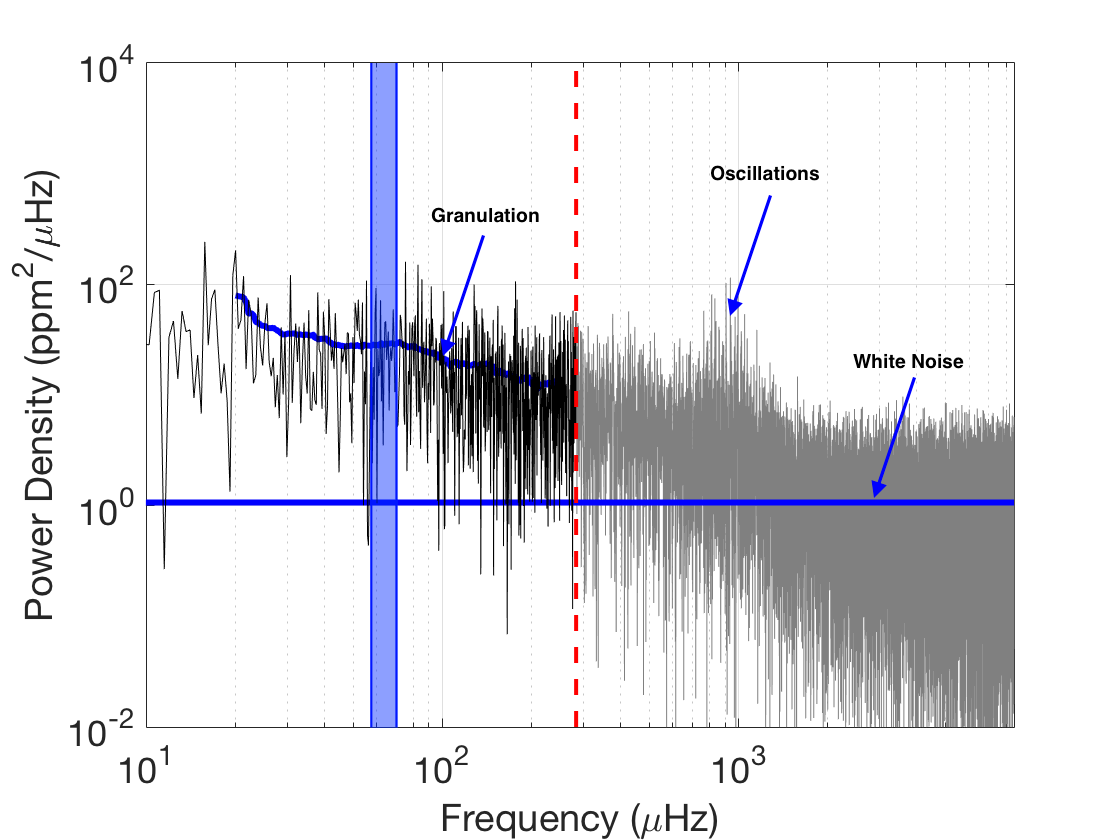}
\caption{Power spectrum of a bright main-sequence star observed by \kepler\
  in short-cadence mode.  The star is KIC~6370489 and the power spectrum is
  based on data from Q1. We see a power excess from solar-like oscillations
  at $\numax = 914\,\muhz$, a granulation background that rises towards
  lower frequencies, and a flat background from white noise. The red dashed
  line indicates the Nyquist frequency for long-cadence observations
  ($283\muhz$).  The blue strip marks a passband centred at $0.08 \numax$
  with fractional bandwidth 20\%.}
\label{fig:example}
\end{center}
\end{figure}

In stars where oscillations cannot be sampled directly because
short-cadence data are not available, we can estimate \logg\ by measuring
the slow variations from granulation.  The strength of these variations
correlates directly with \numax\
\citep{Kjeldsen2011,Chaplin2011-predicting,Mathur2011,Hekker2012,Samadi2013-scaling,Kallinger2014,Bugnet2017,deAssisPeralta2018,Yu2018},
and can therefore provide a way to infer \logg\ (provided \teff\ is known).
This correlation arises because the same underlying mechanism, namely
convection, is responsible for both the oscillations and the granulation.
Note that the photometric variations from granulation behave like $1/f$
noise in that they depend on the timescale on which they are measured.  In
the Fourier power spectrum, this is seen as a background that rises towards
low frequencies (see the example in Fig.~\ref{fig:example}).

%---------------------
\subsection{Short-cadence benchmark sample}\label{sec:short_cadence}

As a first step, we established an empirical relationship between granulation power, \pgran, and the frequency of maximum oscillation power, \numax. To do this, we used the same benchmark sample of 542 \kepler\ stars from
\citet{Huber2011} that was used by \citet{Bastien2013} to calibrate their
relation.  These stars have \numax\ measurements from short-cadence data, with uncertainties derived as described by \citet{Huber2011}.  We analysed the long-cadence data for these stars, as follows:
\begin{enumerate}

\item we downloaded Simple Aperture Photometry (SAP) from MAST for all
  available quarters, and treated each quarter separately;

\item we removed points affected by spacecraft safe-mode events, those with
  quality flag $> 0$, and those with contamination flags $<0.95$;

\item we clipped outliers that were more than 4 standard deviations away
  from the local mean, where the latter was calculated over a 90-minute
  moving window;

\item we applied a high-pass filter \citep{Savitzky1964} to remove very
  slow variations arising from stellar activity, instrumental noise and other
  sources, which could otherwise leak to higher frequencies.  We adapted
  this filter to each star \citep[see also][]{Kallinger2016}, with the cut-off set to
  eliminate power at frequencies below $0.02 \numax$.

\item we calculated the Fourier power spectrum for each quarter up to the
   long-cadence Nyquist frequency ($284\muhz$).  We multiplied by the
   total duration of the observations in order to convert to power density
   (in ${\rm ppm}^2/\muhz$), allowing us to measure the
   granulation background and the white noise level \citep[e.g.,][]{Kjeldsen1999}. 
   
\item we measured the granulation background in the power density spectrum in a bandpass with fractional width 20\% centred at $0.08\numax$, as shown in Fig.~\ref{fig:example}. Our choice of parameters for this bandpass is explained in Sec.~\ref{sec:determine_bandpass}.

\end{enumerate}
We measured the granulation power for each star in this way for every quarter of \kepler\ data (except Q0, the first ten days of observation).  We took the median of these
measurements as our estimate of the granulation power and the standard
deviation over all quarters as its uncertainty.  From this measured
granulation power we subtracted the appropriate white noise correction, as described in the next section.

%-------------------
\subsection{White-noise correction}\label{sec:white_noise}

The \kepler\ light curves include Poisson noise from photon-counting
statistics.  In the Fourier transform this appears as a frequency-independent (flat) noise that is greater for fainter stars.  It is important to correct
for this white noise, since otherwise the granulation power will be
overestimated.  

Measuring the white noise directly requires access to high frequencies,
which is only possible with short-cadence data (see
Fig.~\ref{fig:example}).  We have used a sample of $\sim$2100 stars for
which short-cadence observations are available to construct a calibration
curve of white noise as a function of \kepler\ magnitude,~\kp\ \citep[see
also][]{Gilliland2010}.  This is a much larger group of stars than the benchmark sample introduced in Sec.~\ref{sec:short_cadence} and includes many, especially at the faint end, for which oscillations were not detected.  We calculated Fourier spectra for individual quarters of
\kepler\ data, as described in Sec.~\ref{sec:short_cadence}, and measured the average power density in the region 6000--8000\,\muhz.  This is above the range of oscillation frequencies in
all stars and so provides a good estimate of the white noise.

\begin{figure}
\begin{center}
\includegraphics[width=0.49\textwidth]{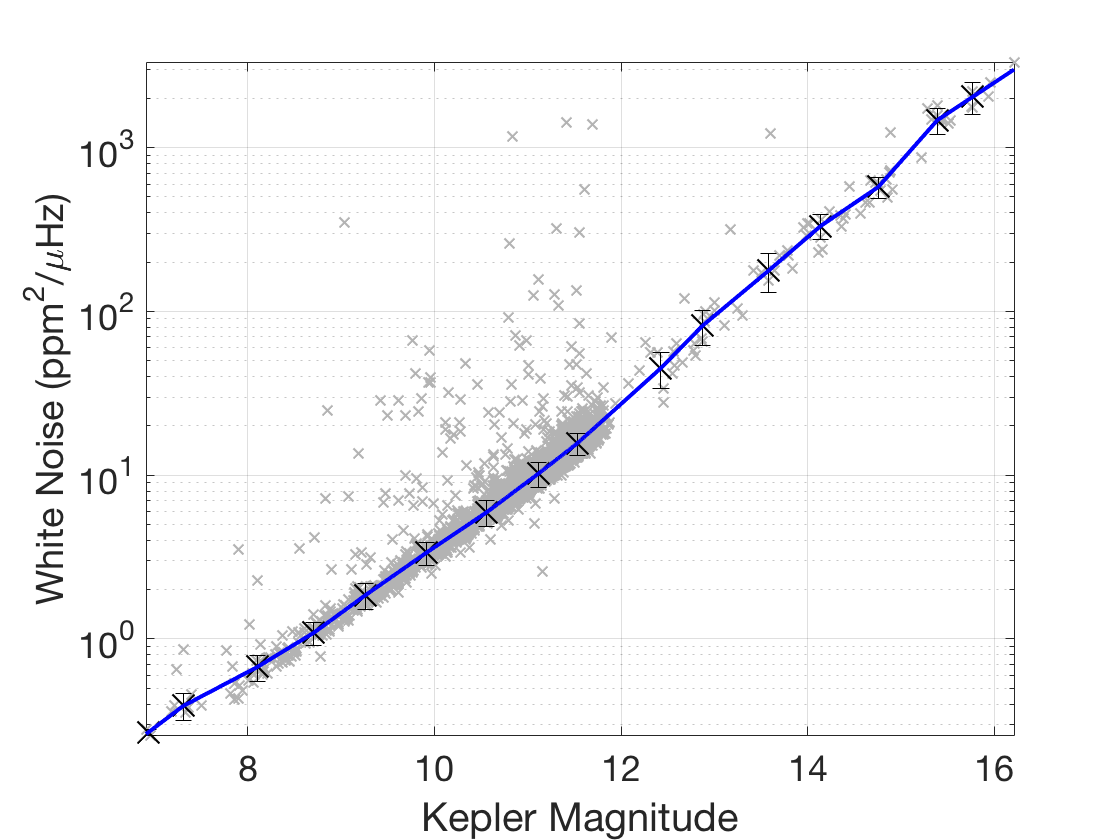}
\caption{Calibration of white noise level as a function of \kepler\
  magnitude. The grey points are measurements from power density spectra of
  $\sim$2100 stars observed in short-cadence mode.  The black points and
  error bars show the median and scatter of points in bins of width
  0.6\,mag.  The blue line connects these points into a piecewise-linear
  calibration curve.}
\label{fig:white_noise}
\end{center}
\end{figure}

The measurements of white noise are shown in Fig.~\ref{fig:white_noise}.
Most of the stars are bright ($\kp < 12$) but there are enough fainter
stars to define the white-noise level accurately down to $\kp = 16$.  Our
calibration curve as a function of \kp\ (blue line) is based on the medians
in bins of width 0.6 magnitudes.  The error bars show the uncertainties on
the correction, which are calculated as the median absolute deviation
within each bin.  We subtracted this white-noise level from the granulation
power measured for each star in the benchmark sample (Sec.~\ref{sec:short_cadence}), and
also for the main sample (Sec.~\ref{sec:results}). The uncertainty in the white noise correction 
was added in quadrature to the quarter-to-quarter uncertainty estimate of granulation power described in the next section.

%------------------------------------------------------------------------------

\begin{figure}
\begin{center}
\resizebox{\hsize}{!}{\includegraphics{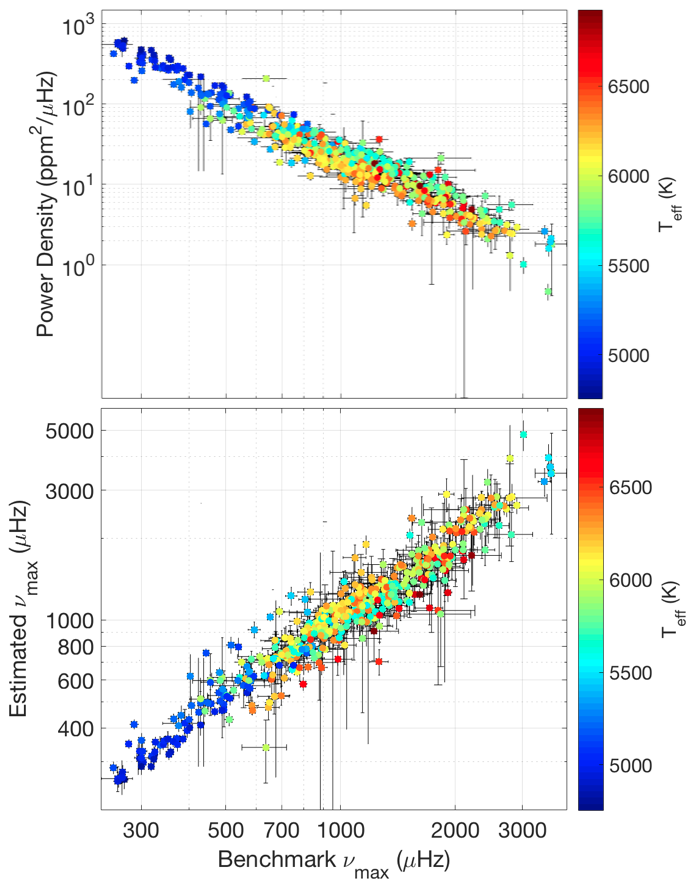}}
\caption{Granulation power (upper) and the inferred value of \numax\ using Eq.~\ref{eq:numax_power} (lower) for the benchmark sample, plotted against \numax\ measured by \citet{Huber2011} from short-cadence data. Colour represents~\teff\ \citep{Mathur2017}. }
\label{fig:benchmark}
\end{center}
\end{figure}

\subsection{Calibrating the relation} \label{sec:calibration}

The upper panel of Fig.~\ref{fig:benchmark} shows our measured granulation
power density, \pgran, for each star in the benchmark sample after correcting for white noise.  These values
are plotted against \numax\ for each star, as measured by \citet{Huber2011}  
from the oscillations themselves using short-cadence data.  We confirm the
strong correlation that was previously found from \kepler\ data
\citep{Chaplin2011-predicting,Huber2011,Samadi2013-scaling,Kallinger2014}.
However, we also see a clear dependence on effective temperature,
indicating that we should include \teff\ in our fit.

We calculated the fit as a linear function relating $\log \numax$, $\teff$
and $\log\pgran$ as follows:
\begin{align} \label{eq:numax_power}\begin{split}
\log (\numax /\muhz)  = -0.4861 \log (\pgran/({\rm ppm}^2/\muhz)) - \\
\teff/(11022\,{\rm K}) + 4.197.\end{split}
\end{align}
The lower panel of Fig.~\ref{fig:benchmark} shows the relation between \numax\
estimated from this fit and the measured benchmark values. The comparison show good agreement,  with systematic variations below the 5\% level. 
Converting \numax\ to \logg\ via Eq.~\ref{eq:numax.scaling} gives \logg\
values for the benchmark sample with a scatter of $\sim$0.05 \,dex.

%------------------------------------------------------------

\subsection{Determining the bandpass} \label{sec:determine_bandpass}

As noted in Sec.~\ref{sec:short_cadence}, we measured the granulation background in the power spectrum of each star adaptively at a fixed fraction of \numax.  The frequency of this bandpass
should be low enough that it falls below the long-cadence Nyquist frequency
for all stars of interest, and high enough that it is unaffected by the
high-pass filter described in Sec.~\ref{sec:short_cadence}.  We tested bandpasses with centres in the range from $0.04\numax$ to $0.10\numax$.  We also tested different values
for the fractional width of the bandpass, from 10\% to 50\%.

%\new{[Omit next sentence: Combining this with the effective temperature of each star from the latest version of the Kepler Stellar Properties Catalog \citep{Mathur2017} in Eq.~\ref{eq:numax.scaling} allowed us to estimate \numax\ for all benchmark stars.]}
%
\new{In order to choose the optimal bandpass, we sought to minimize the scatter between our derived values of \numax\ and those measured by \citet{Huber2011}, as plotted in the lower panel of Fig.~\ref{fig:benchmark}.}  At the same time, we also aimed to avoid excluding stars for which this bandpass would lie above the Nyquist frequency for long-cadence data.  We settled on a bandpass with fractional width 20\% centred at $0.08\numax$, as shown in
Fig.~\ref{fig:example}. 

%----------------------------------------------------------------

\section{Results and Discussion} 

\subsection{The catalogue of surface gravities} \label{sec:results}

We applied our method to measure \logg\ for the same sample of long-cadence stars that was
analysed by B16.  This comprises 28,715 stars with  $\kp\ < 13.5$, the majority of which
have $\kp > 12$.  We excluded 262 stars that are listed in the Kepler Eclipsing Binary Catalog by \citet{Kirk2016}, since the eclipses have harmonics in the power spectrum that interfere with measuring the granulation background.

Our method requires knowledge of \numax\ in order to carry out the high-pass filtering and also to set the bandpass in which \pgran\ is measured (Sec.~\ref{sec:short_cadence}).  We therefore adopted an iterative approach, with the initial estimate of \numax\ being calculated from values of \logg\ and \teff\ in the Kepler Input Catalog (KIC) \citep{Brown2011}.  In practice, we found that the choice of
this initial \numax\ did not significantly affect the final result in most cases.

\begin{figure}
\begin{center}
\includegraphics[width=0.49\textwidth,trim={0cm 1cm 1cm 1cm}]{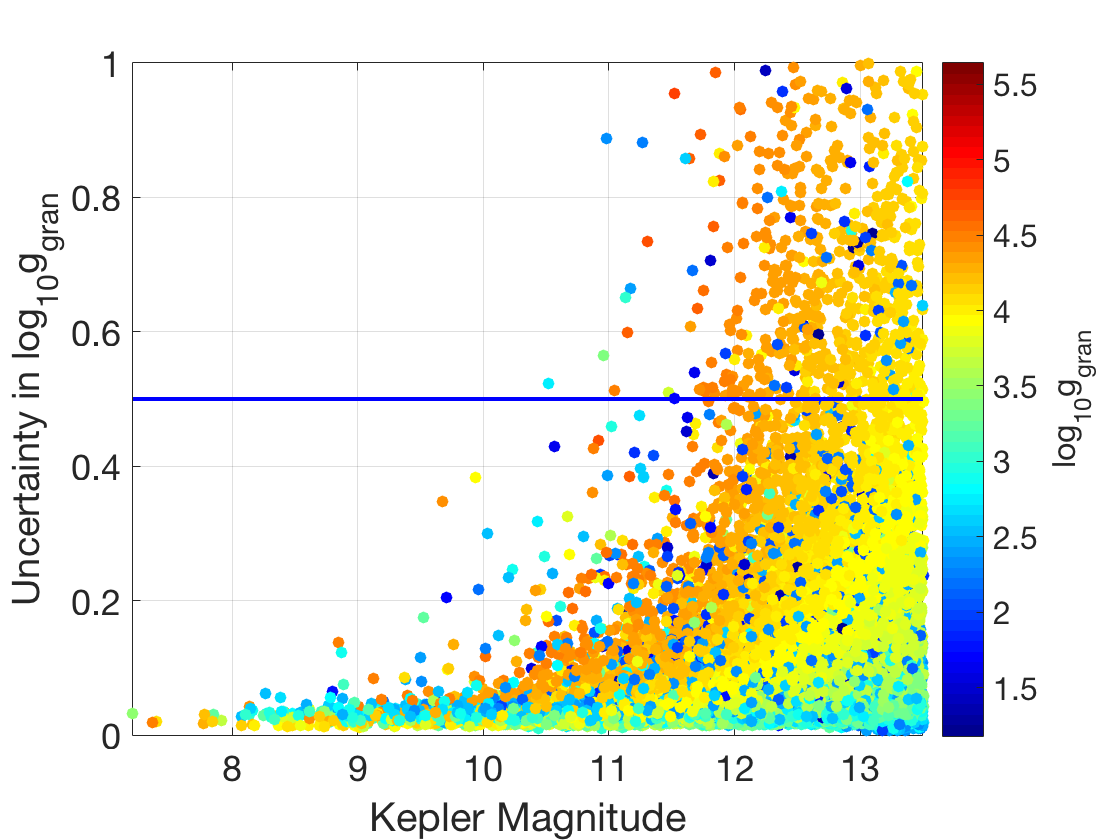}
\caption{The uncertainty in our \logg\ estimates plotted against
\kp. Colour represents our \logg\ estimates. The horizontal blue line shows
the cut we made at an uncertainty of 0.5 dex.}
\label{fig:bastien_sampleuncertainty}
\end{center}
\end{figure}

For 4181 giants with $\numax < 100\,\muhz$, the process did not converge
after three iterations and those stars were excluded from the sample.  For a further
818 stars, mostly cool dwarfs, the granulation power was too close to the
white noise level to be measured.  The remaining sample contained 23454
stars.  Figure \ref{fig:bastien_sampleuncertainty} shows their \logg\
uncertainties as a function of \kepler\ magnitude.  We see that many
have large uncertainties, mainly faint dwarfs, indicating the
difficulty of measuring their granulation signal above the white noise.
Our final catalogue, for which estimates of \pgran, \numax\ and \logg\ are
listed in Table \ref{tab:fulltable}, comprises 15109 stars having
uncertainties in \logg\ smaller than 0.5 dex.

\begin{table*}
\caption{Estimates of granulation power, \numax\ and \logg\ for 15,109
stars using \kepler\ long-cadence data.  Effective temperatures are taken
from \citet{Mathur2017}. (This table is available in its entirety in a
machine-readable form in the online journal. A portion is shown here for
guidance regarding its form and content.)  }

\begin{center}
\begin{tabular}{cccccc}\hline
KIC     & Kp     & \teff\ (K) & \pgran\ (${\rm ppm}^2/\muhz$) & \numax\ (\muhz) & \logg \\\hline
1025494 & 11.822 & 6122(172) &    32.20(5.12)    & 810.7(91.8)  & 3.87(0.05)\\
1026084 & 12.136 & 5072(166) & 21357.21(3217.25) &  42.9(4.6)   & 2.55(0.05)\\
1026255 & 12.509 & 7050(214) &   151.11(17.57)   & 315.0(31.9)  & 3.49(0.04)\\
1026475 & 11.872 & 6611(189) &    18.78(6.18)    & 951.3(189.8) & 3.96(0.09)\\
1026669 & 12.304 & 6293(193) &    21.24(9.15)    & 957.6(239.0) & 3.95(0.11)\\ \hline
\end{tabular}
\end{center}
\label{tab:fulltable}
\end{table*}

It is worth noting that the granulation power in red giants has a slight metallicity dependence, in the sense that metal-rich stars have stronger granulation \citep{Corsaro2017,Yu2018}. If such a relation applied to the main-sequence and subgiant stars studied here, it would affect the values of \logg\ derived from measuring granulation power.

\subsection{Comparison with Spectroscopy}

\begin{figure}
\begin{center}
\includegraphics[width = 0.45\textwidth,trim={0cm 0cm 1cm 1cm}]{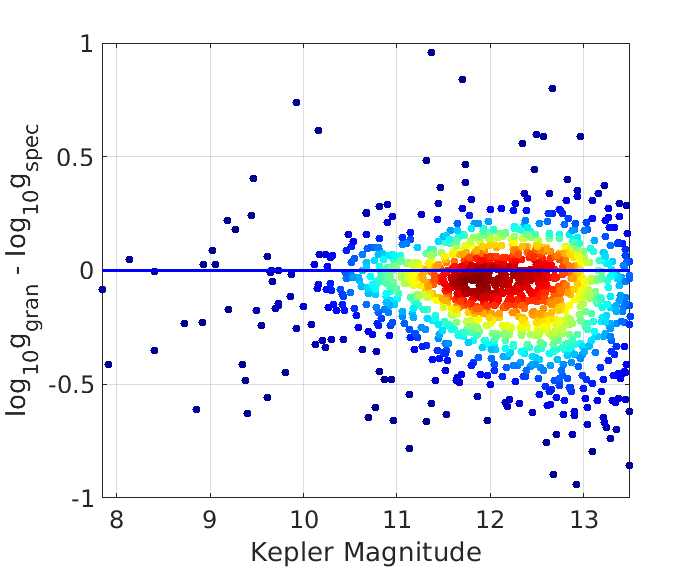}
\caption{Difference in \logg\ between our estimates and those from spectroscopy, plotted against \kp. Colours
represents the density of stars in each region.}
\label{fig:comparison_spectroscopy}
\end{center}
\end{figure}

We compared our estimates with those obtained from
spectroscopy in Fig.~\ref{fig:comparison_spectroscopy}. We adopted spectroscopic 
parameters for 3017 stars from the Kepler Stellar Properties Catalog \citep{Mathur2017}, which primarily contains values from LAMOST \citep[2075 sources,][]{Luo2015,DeCat2015}, APOGEE \citep[588 sources,][]{Alam2015} and \citet{Buchhave2014}. The comparison generally shows good agreement  and no trend with \kp\ but there is a small systematic offset of $\sim$0.05\,dex.

Spectroscopic estimates of \logg\ should generally be unaffected by distance or interstellar reddening. On the other hand, estimating \logg\ from photometric fluctuations is potentially very sensitive to the way that white photon noise is accounted for. 
Therefore, the fact that we do not see a systematic trend with \kp\ in Fig.~\ref{fig:comparison_spectroscopy}
gives us confidence that our white-noise correction is effective. For comparison, such a trend is seen in Fig.~1 of \citet{Bastien2014}.

%-----------------------------------------------------------------------------

\subsection{Comparison with Gaia Parallaxes}

The recent release of \gaia\ DR2 parallaxes \citep{Brown2018, Lindegren2018} provides another opportunity to test our \logg\ values and to validate granulation-based surface gravities with a larger sample than the benchmark asteroseismic detections from \kepler\ short-cadence data. To do this, we combined the \logg\ values from Table~\ref{tab:fulltable} with effective temperatures and metallicities from \citet{Mathur2017} to calculate luminosities from isochrones using the software package \texttt{isoclassify} \citep{huber17}. We restricted the analysis to the $\sim$\,13,400 stars in our sample with \logg\ uncertainties below 0.3\,dex.

\begin{figure*}
\begin{center}
\resizebox{\hsize}{!}{\includegraphics{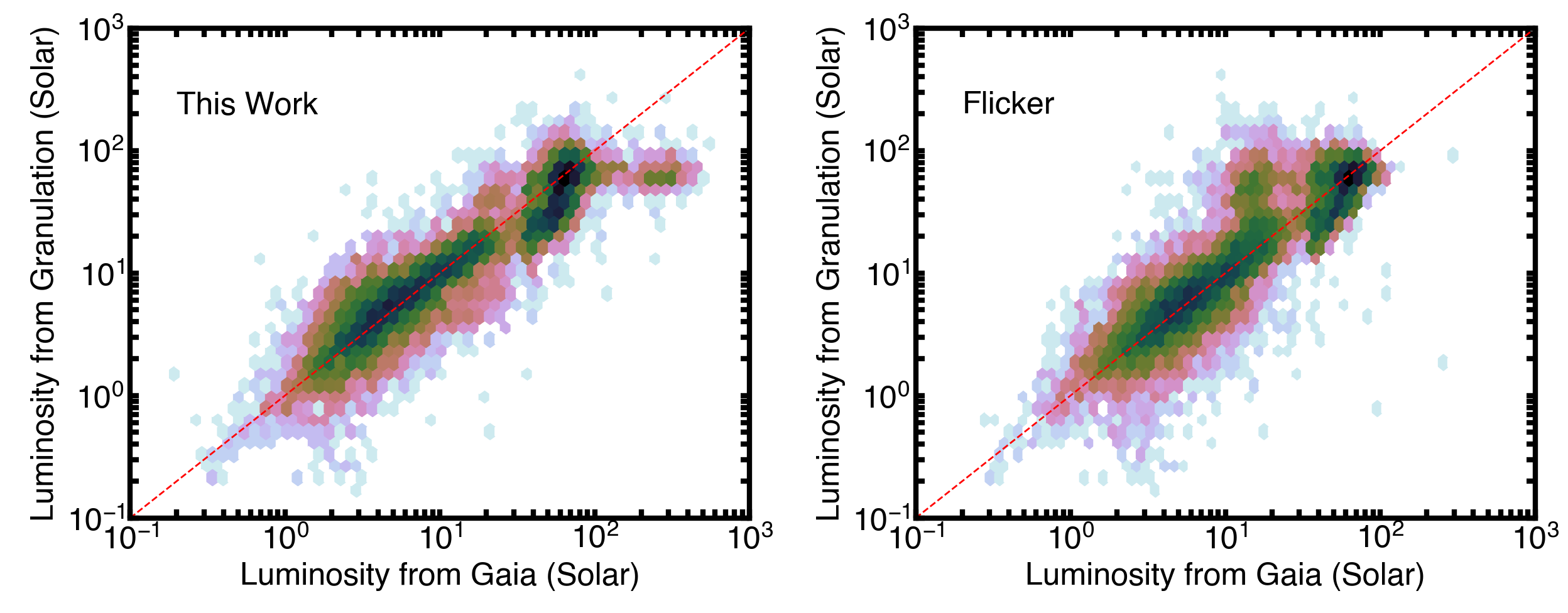}}
\caption{Comparison of luminosities derived from Gaia DR2 parallaxes from \citet{berger18} with luminosities derived the granulation-based \logg\ values in our work (left panel) and \logg\ values from flicker (right panel, B16). Only stars with \logg\ values measured to better than 0.3\,dex with our method are shown in both panels. Colors show logarithmic number density, and the red line marks equality.}
\label{fig:gaia}
\end{center}
\end{figure*}

The left panel of Fig.~\ref{fig:gaia} compares our luminosities to those derived from \gaia\ parallaxes by \citet{berger18}. We see good agreement over three orders of magnitude, but a systematic discrepancy for high-luminosity red giants ($\gtrsim 100 \lsun$, $\logg < 2.3$), for which our \logg\ values are systematically too small. We suspect that this difference is due to the extrapolation of our calibration, which is mostly based on main-sequence stars and subgiants (Fig.~\ref{fig:benchmark}). Excluding these high-luminosity giants, we find a residual scatter of 40\%, which is roughly consistent with the typical uncertainties on the granulation-derived luminosities (which includes uncertainties on \teff, \logg\ and \feh). Systematic differences are at the level of 7\% or less in luminosity, which 
\old{translates into an accuracy of} 
\new{corresponds to}
$\sim$\,0.07\,dex in \logg. However, we note that systematic errors in the \teff\ scale may account for part of this. The right panel of Figure \ref{fig:gaia} shows the same comparison but using flicker-derived \logg\ values from B16. As expected, the overall performance of  flicker-derived \logg\ values is similar to ours, with a slightly higher scatter and different systematics for high-luminosity giants. 

Overall, this comparison with \gaia\ validates granulation as a powerful tool to measure \logg\ from light curves with 
\old{significantly better precision than spectroscopy} 
\new{good precision.} 
Combined with radii from \gaia\ parallaxes, this may allow the measurement of masses for a large number of individual stars without using stellar models \citep{stassun18}. However, we caution that systematic differences at the $<0.1$\,dex level due to uncertainties in the \numax\ scaling relations and white noise corrections still need to be carefully quantified.

\section{Conclusion}

We have estimated \logg\ from \kepler\ light curves by measuring the granulation background, similar to the ``flicker'' method by \citet{Bastien2016} but working in the Fourier power spectrum.  We established a calibration for white noise as a function of magnitude (Fig.~\ref{fig:white_noise}) and calibrated the granulation power using the asteroseismic short-cadence sample, demonstrating a precision in \logg\ of about 0.05\,dex (Fig.~\ref{sec:short_cadence}). Applying the method to the sample of 28,000 long-cadence stars studied by \citet{Bastien2016}, we found about half the stars to be too faint for the granulation background to be reliably detected above the white noise.  We have provided an electronic catalogue of \logg\ values (with uncertainties) for about 15,000 stars having \logg\ uncertainties better than 0.5\,dex (Table~\ref{tab:fulltable}).

There is no magnitude-dependent trend in the difference between our estimates and those available from 
spectroscopy, giving us confidence that our white noise correction is effective. We also use \gaia\ DR2 parallaxes to validate that granulation is a powerful method to measure \logg\ from light curves. Our method can also be applied to the large number of light curves collected by K2 and TESS.

\section*{Acknowledgements}

We gratefully acknowledge support from the Australian Research Council, and
from the Danish National Research Foundation (Grant DNRF106) through its
funding for the Stellar Astrophysics Centre (SAC). D.H. acknowledges support by the Australian Research Council's Discovery Projects funding scheme (project number DE140101364) and support by the National Aeronautics and Space Administration under Grant NNX14AB92G issued through the Kepler Participating Scientist Program.  \new{This work has made use of data from the European Space Agency (ESA) mission
{\it Gaia} (\url{https://www.cosmos.esa.int/gaia}), processed by the {\it Gaia}
Data Processing and Analysis Consortium (DPAC,
\url{https://www.cosmos.esa.int/web/gaia/dpac/consortium}). Funding for the DPAC
has been provided by national institutions, in particular the institutions
participating in the {\it Gaia} Multilateral Agreement.} \new{We thank Matthias Ammler-von Eiff and the referee, Gibor Basri, for helpful comments on this paper.} 


\bibliographystyle{mnras}
\bibliography{bibliography}

% Don't change these lines
\bsp	% typesetting comment
\label{lastpage}
\end{document}